\documentclass[10pt]{iopart}
\usepackage{graphicx,iopams,cite}

\usepackage[T1]{fontenc}
\usepackage[latin1]{inputenc}

\usepackage[usenames]{color}
\usepackage[latin1]{inputenc}
\usepackage{ulem}

\newcommand{\eq}[1]{Eq.~(\ref{#1})}
\newcommand{\fig}[1]{Fig.~\ref{#1}}

\newcommand{\sect}[1]{Section~\ref{#1}}
\newcommand{\avg}[1]{\langle #1 \rangle}

\newcommand{\ahum}[1]{``#1''}
\newcommand{\figwidth}{0.95\columnwidth}

\newcommand{\twothirdfigwidth}{0.66\columnwidth}
\newcommand{\beq}{\begin{equation}}
\newcommand{\eeq}{\end{equation}}
\newcommand{\bea}{\begin{eqnarray}}
\newcommand{\eea}{\end{eqnarray}}

\newcommand{\defas}{:=}
\newcommand{\tc}{T_{\rm c}}
\newcommand{\etapr}{\eta_{\rm p}^{\rm r}}
\newcommand{\etaprcr}{\eta_{\rm p,cr}^{\rm r}}
\newcommand{\zc}{z_{\rm c}}
\newcommand{\zacx}{z_{\rm cx}}
\newcommand{\uF}{u_L}
\newcommand{\rmin}{R_{\rm min}(\theta, \nu)}

\begin{document}

\title[Fluids with quenched disorder]{Fluids with quenched disorder: Scaling of 
the free energy barrier near critical points}

\author{T.~Fischer and R.~L.~C.~Vink}

\address{Institute of Theoretical Physics, Georg-August-Universit\"at
G\"ottingen, Friedrich-Hund-Platz~1, D-37077 G\"ottingen, Germany}

\begin{abstract} In the context of Monte Carlo simulations, the analysis of the 
probability distribution $P_L(m)$ of the order parameter $m$, as obtained in 
simulation boxes of finite linear extension~$L$, allows for an easy estimation 
of the location of the critical point and the critical exponents. For Ising-like 
systems without quenched disorder, $P_L(m)$ becomes scale invariant at the 
critical point, where it assumes a characteristic bimodal shape featuring two 
overlapping peaks. In particular, the ratio between the value of $P_L(m)$ at the 
peaks ($P_{L, \rm max}$) and the value at the minimum in-between ($P_{L, \rm 
min}$) becomes $L$-independent at criticality. However, for Ising-like systems 
with quenched random fields, we argue that instead $\Delta F_L \defas \ln \left( 
P_{L, \rm max} / P_{L, \rm min} \right) \propto L^\theta$ should be observed, 
where $\theta>0$ is the \ahum{violation of hyperscaling} exponent. Since 
$\theta$ is substantially non-zero, the scaling of $\Delta F_L$ with system size 
should be easily detectable in simulations. For two fluid models with quenched 
disorder, $\Delta F_L$ versus $L$ was measured, and the expected scaling was 
confirmed. This provides further evidence that fluids with quenched disorder 
belong to the universality class of the random-field Ising model. \end{abstract}

\maketitle

\section{Introduction}

It was postulated by de~Gennes that liquid-gas type transitions in fluids 
confined to quenched random porous media belong to the universality class of the 
random field Ising model (RFIM) \cite{gennes:1984}. The unambiguous experimental 
verification of this conjecture remains elusive to this day 
\cite{citeulike:5845715}, but there is growing numerical evidence from computer 
simulations \cite{citeulike:3523819, citeulike:4000781, citeulike:7690917} of 
quenched-annealed mixtures \cite{madden.glandt:1988}. In quenched-annealed 
mixtures, a fluid of mobile particles is confined to a configuration of immobile 
(quenched) particles. The quenched particles induce a random field provided (1) 
their positions are sufficiently random and (2) they display a preferred 
affinity to one of the phases formed by the mobile particles. When either one of 
these conditions is not met, the conjecture of de~Gennes does not apply 
\cite{citeulike:3773391, citeulike:5845715}.

Computer simulations of quenched-annealed mixtures remain complicated. One 
problem is that for systems exhibiting quenched disorder an additional average 
over many samples drawn from the distribution characterizing the disorder needs 
to be taken. In addition to the thermal averaging, a disorder averaging over the 
different samples must be taken, and hence the computational effort is orders of 
magnitudes larger. The convergence of disorder averaged quantities with the 
number of samples is typically slow (random field systems at criticality do not 
self average \cite{citeulike:7465852, citeulike:6672115, aharony.harris:1996, 
malakis.fytas:2006, wiseman.domany:1995, citeulike:6348195}) and so the disorder 
averaging must comprise many samples. A second problem is that the critical 
exponents of the three-dimensional RFIM are not known very precisely. In 
particular, estimates for the correlation length exponent $\nu$ range from $1.1$ 
to $2.25$ \cite{nattermann:1998}. Due to the large uncertainty in $\nu$, 
standard finite size scaling (FSS) of simulation data is problematic. The only 
critical exponent of the RFIM on which there is some consensus, is the 
\ahum{violation of hyperscaling} exponent $\theta$. In pure systems, meaning 
systems without quenched disorder, it holds that $\theta=0$, but in the RFIM 
$\theta \sim 1.5$, i.e.~distinctly non-zero. In order to detect RFIM 
universality, it follows that FSS of a quantity \ahum{somehow involving} the 
violation of hyperscaling exponent $\theta$ is a promising approach. In this 
paper, we present one such strategy, based on the free energy cost of interface 
formation. In the RFIM at its critical point, the barrier diverges with the 
system size as 
\beq\label{eq:df}
 \Delta F_L \propto L^\theta, 
\eeq
where $L$ is the lateral extension of the simulation box 
\cite{citeulike:7690917}. In the pure Ising model where $\theta=0$, $\Delta F_L$ 
is $L$-independent at criticality \cite{physrevlett.65.137, binder:1981, 
citeulike:7690917}. Since $\theta$ is large in the RFIM, the divergence of 
$\Delta F_L$ should be easily detectable in random field systems.

The outline of this paper is as follows. We first describe how the scaling of 
the free energy barrier at criticality can be derived and exploited in the 
context of FSS. We then illustrate the approach for the RFIM, the 
Widom-Rowlinson (WR) model \cite{widom.rowlinson:1970} with quenched obstacles, 
and the Asakura-Oosawa (AO) model \cite{asakura.oosawa:1954, vrij:1976} of 
colloid-polymer mixtures inside a porous medium. In all cases $\theta>0$ is 
observed, but the precise value varies. For the RFIM and WR mixture $\theta \sim 
1.5$ is obtained, while colloid-polymer mixtures yield a somewhat lower value 
$\theta \sim 1.0$; possible origins for this discrepancy are discussed in 
\sect{con}.

\section{Finite size scaling of the interface free energy}

\subsection{theoretical background}

We assume that the RFIM in $d=3$ dimensions undergoes a continuous phase 
transition, at critical temperature $\tc$, from a disordered phase at high 
temperature, to an ordered phase with finite magnetization at low temperature. The 
existence of a transition at nonzero temperature was controversial until a proof 
for the existence of a spontaneous magnetization settled this issue 
\cite{imbrie:1984}; rigorous results on the order of the transition remain 
elusive, but numerical studies \cite{rieger:1995, newman.barkema:1996} indicate 
that the transition is continuous. Following convention, we define the relative 
distance from the critical point as
\beq\label{eq:t}
 t \defas T/\tc-1. 
\eeq
In the vicinity of the critical point, $t=0$, the correlation length diverges as 
a power law
\beq\label{eq:cl}
 \xi \propto |t|^{-\nu}, 
\eeq
while in the ordered phase a finite interfacial tension (excess free 
energy per unit of interface) develops
\beq
 \sigma \propto |t|^{2-\alpha-\nu} \quad (t<0),
\eeq
where $\alpha$ is the critical exponent of the specific heat. Near the critical 
point, the correlation length is the only relevant length scale, and so a simple 
dimensional argument implies that the total interface free energy must scale as
\beq
 \Delta F_\xi \propto \sigma \, \xi^{d-1} \propto 
 |t|^{2-\alpha-\nu} \xi^{d-1} \propto
 \xi^{(\alpha + d\nu - 2)/\nu},
\eeq
where in the last step $t$ was eliminated using \eq{eq:cl}. If we now use the 
FSS \ahum{Ansatz} $\xi \propto L$, the above equation reduces to
\beq\label{eq:dfl}
 \Delta F_L \propto L^{(\alpha + d\nu - 2)/\nu},
\eeq
which describes the scaling of the interface excess free energy in a finite 
simulation box of lateral extension $L$ in the vicinity of $\tc$. This equation 
applies to both the pure Ising model and the RFIM. The hallmark of RFIM 
universality is the modified hyperscaling relation \cite{schwartz:1985, 
schwartz.gofman.ea:1991, gofman.adler.ea:1993}
\beq
 2 - \alpha = \nu (d - \theta),
\eeq
which upon substitution in \eq{eq:dfl} yields
\beq \label{eq:dflt}
 \Delta F_L \propto L^\theta,
\eeq
and so we have derived \eq{eq:df}. In the pure Ising model $\theta=0$, in which 
case the barrier does not depend on $L$ at the critical point (as is well known 
\cite{physrevlett.65.137}). In the RFIM $\theta \sim 1.5$, implying a strong 
divergence with system size.

\subsection{order parameter distributions}

We now explain how the barrier $\Delta F_L$ is measured in computer simulations 
of liquid-gas type transitions. For simplicity, we first consider a single 
component fluid in the absence of quenched disorder. The key quantity is the 
order parameter distribution (OPD)
\beq \label{eq:opd}
\begin{array}{lcr}
 P_L(\rho|T,\mu) & & \mbox{(single component fluid)}, 
\end{array}
\eeq
defined as the probability to observe a state with particle density $\rho$. The 
OPD is measured in the grand canonical ensemble, at fixed temperature $T$ and 
chemical potential $\mu$, using a periodic $L \times L \times L$ simulation box 
\cite{frenkel.smit:2001, vink.horbach:2004*1}. Below the critical temperature 
$\tc$ and at the coexistence chemical potential $\mu_{\rm cx}$, the OPD becomes 
bimodal; the peak at low (high) density corresponds to the gas (liquid) phase. 
The coexistence chemical potential is obtained by maximizing the derivative of 
the average density with respect to $\mu$
\beq\label{eq:cx}
 \mu_{\rm cx} : \partial \avg{\rho} / \partial \mu \to \mbox{max},
\eeq
with $\avg{\rho} = \int \rho \, P_L(\rho|T,\mu) \, d\rho$. Note that other 
choices to define $\mu_{\rm cx}$ are possible also 
\cite{orkoulas.fisher.ea:2001}, most notably the \ahum{equal-weight} rule 
\cite{borgs.kappler:1992}, but \eq{eq:cx} has the advantage that it remains 
well-defined also when the peaks in the OPD overlap. The free energy barrier 
$\Delta F_L$ is encoded in the logarithm of the OPD: it corresponds to the 
average peak height of $\ln P_L(\rho)$, measured from the minimum 
\ahum{in-between} the peaks \cite{binder:1982}. 

\begin{figure}
\begin{center}
\begin{tabular}{ccc}
\includegraphics[width=0.45\columnwidth]{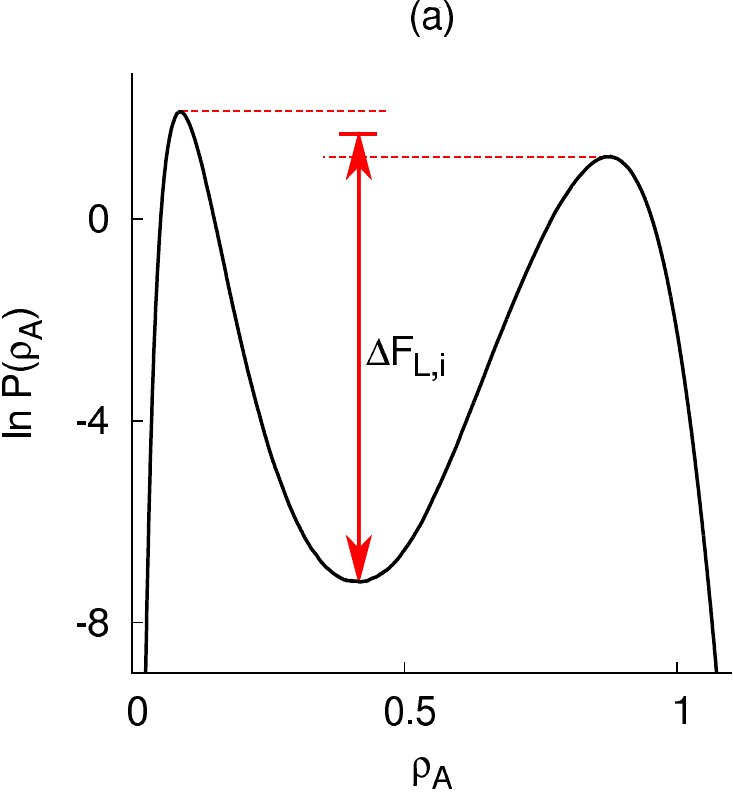} & \hspace{0.05\columnwidth} &
\includegraphics[width=0.45\columnwidth]{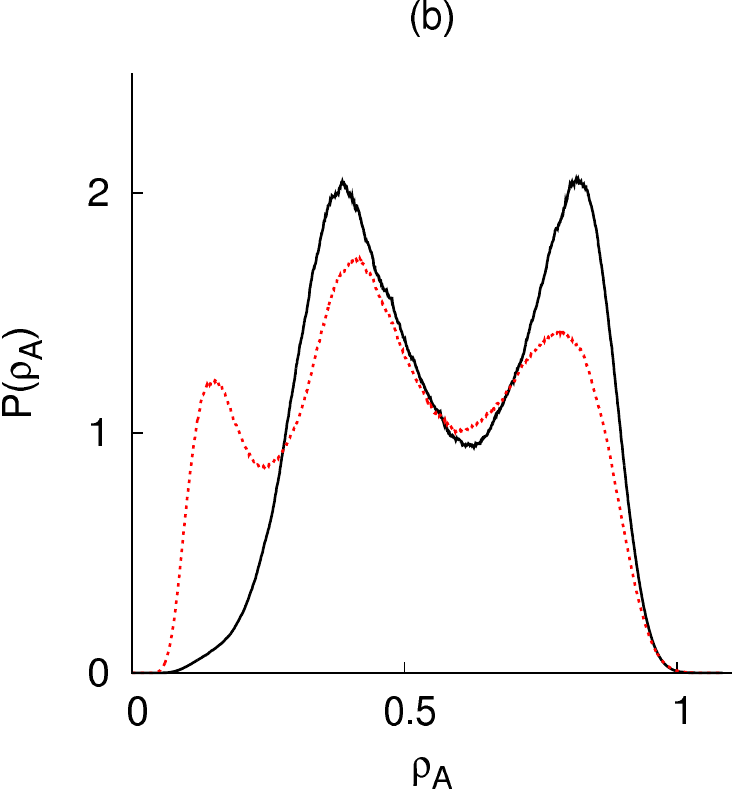}
\end{tabular}
\caption{\label{fig1} (a) Example OPD for the WR mixture (with quenched 
obstacles) obtained for $z_B=1.2$, $L=12$, and $z_A$ chosen according to 
\eq{eq:cx}. Note that the natural logarithm of the distribution is shown. The 
barrier $\Delta F_{L,i}$ corresponds to the average peak height (vertical 
arrow). (b) Example of two \ahum{problematic} OPDs, again for $z_B=1.2$, $L=12$, 
and $z_A$ tuned according to \eq{eq:cx}. These distributions feature a peak at 
intermediate density, which does not correspond to a gas or liquid phase. The 
intermediate peak either replaces the gas or liquid peak, in which case the 
resulting distribution remains bimodal (solid curve), or it appears as a third 
peak (dashed curve). A free energy barrier corresponding to coexisting gas and 
liquid phases cannot be extracted from these distributions.}
\end{center}
\end{figure}

In athermal binary mixtures with mobile species $A$ and $B$ (but still without 
disorder), the analogue of \eq{eq:opd} becomes
\beq\label{eq:opdab}
 \begin{array}{lcr}
 P_L(\rho_A|z_A,z_B) & & \mbox{(athermal binary mixture)},
 \end{array}
\eeq
where $\rho_A$ is the density of the $A$ particles, and where $z_A$ and $z_B$ 
are the fugacities of the $A$ and $B$ particles, respectively. By athermal we 
mean that the particle interactions are either hard-core or ideal, as is the 
case in the WR and AO models. Those mixtures correspond to a single component 
fluid if one identifies $\rho \leftrightarrow \rho_A$, $T \leftrightarrow 
1/z_B$, and $\mu \leftrightarrow \ln z_A$. The transition between a \ahum{gas
phase} with low $\rho_A$ and a \ahum{liquid phase} with high $\rho_A$ is now 
driven by the fugacities. It occurs on the coexistence curve $z_A = 
\zacx (z_B)$, provided the fugacity of the $B$ particles exceeds the critical 
fugacity $z_B > \zc$. On the coexistence curve, which again can be found using 
\eq{eq:cx}, the OPD becomes bimodal and allows the extraction of a free energy 
barrier $\Delta F_L$, see \fig{fig1}(a).

\begin{figure}
\begin{center}
\begin{tabular}{ccc}
\includegraphics[width=0.25\columnwidth]{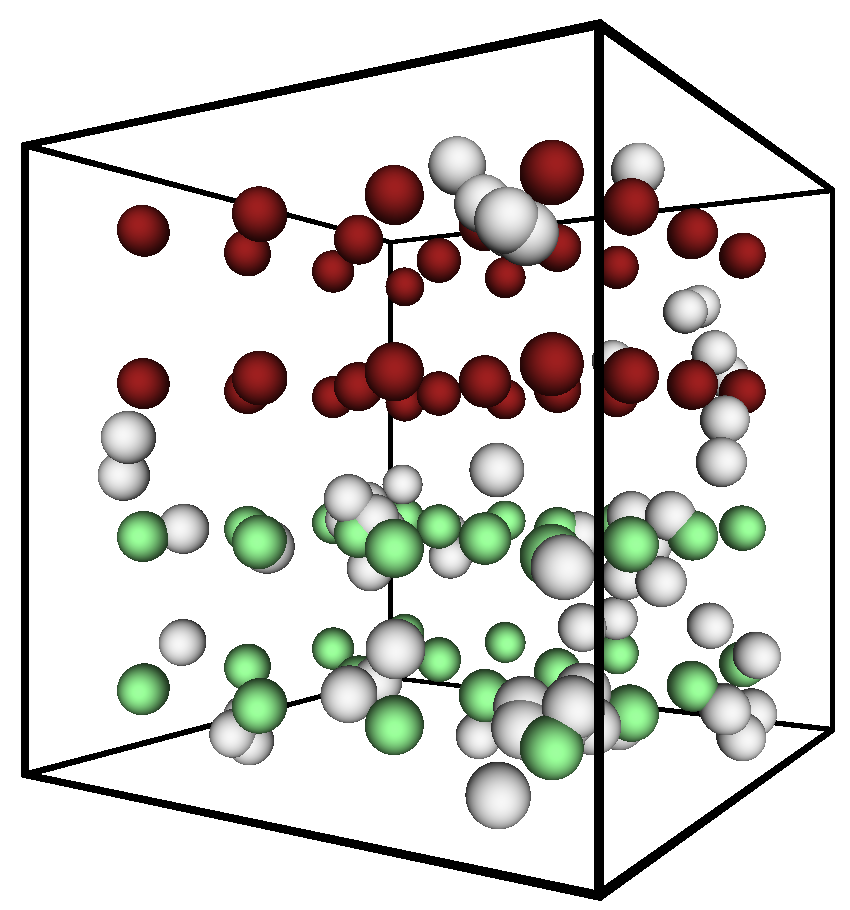} &
\includegraphics[width=0.25\columnwidth]{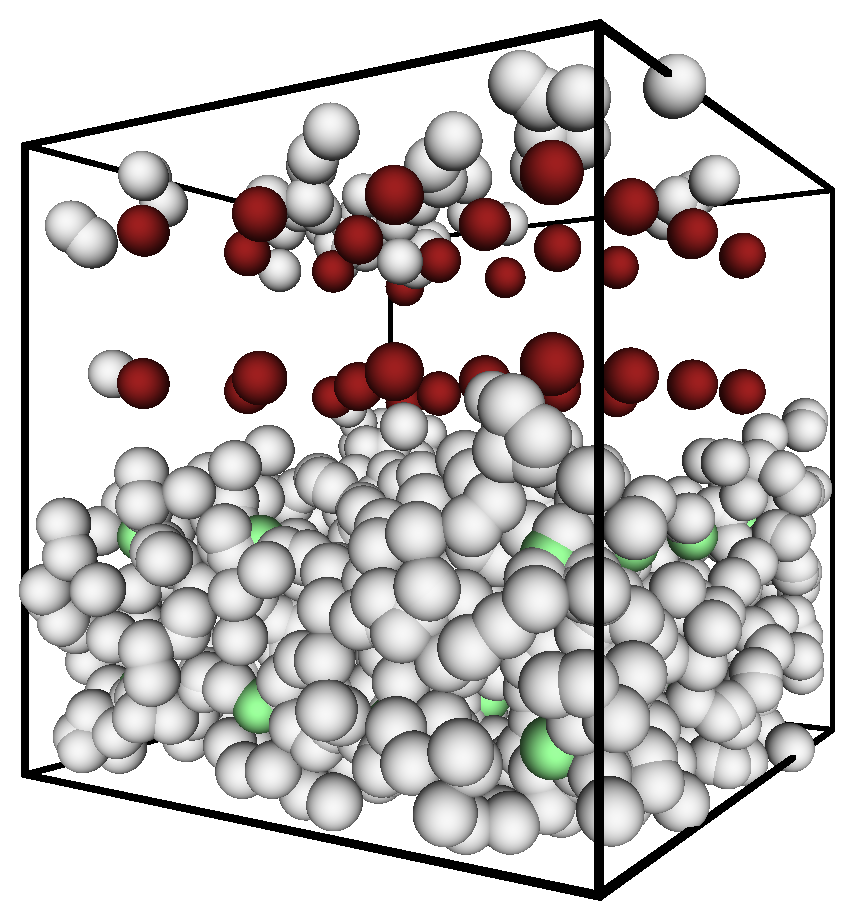} &
\includegraphics[width=0.25\columnwidth]{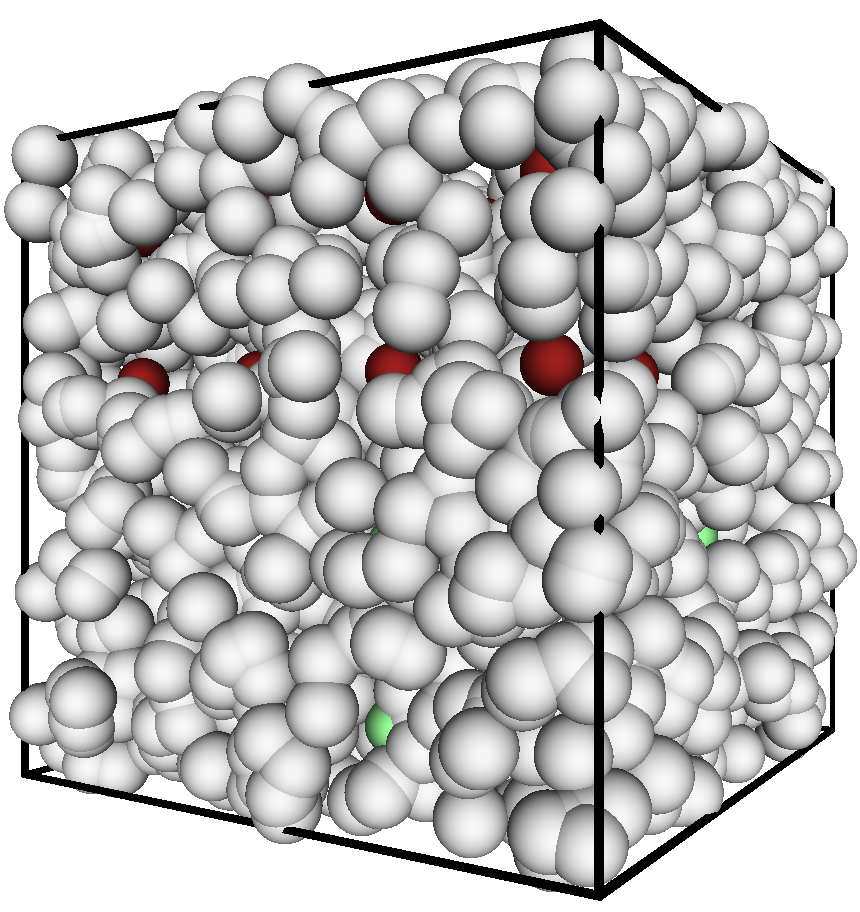}
\end{tabular} \\
\vspace{0.05\columnwidth}
\includegraphics[width=\twothirdfigwidth]{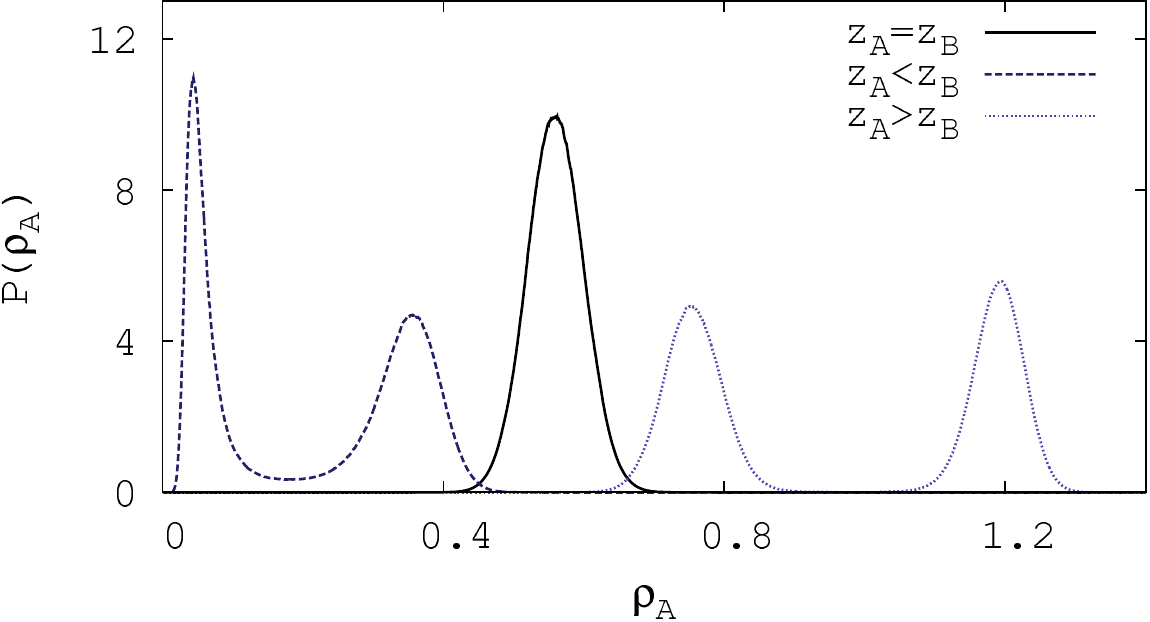}

\caption{\label{fig:snapshots} Illustration of how the \ahum{problematic} OPDs 
in \fig{fig1}(b) might arise. We provide data for the WR model (particle species 
$A$ and $B$) with a configuration of quenched obstacles explicitly tailored to 
feature a peak in the OPD at intermediate density. Instead of distributing the 
obstacles randomly, they are placed on a regular grid; the obstacles in the 
lower half of the grid favor $A$ particles, those in the upper half $B$ 
particles. This configuration yields three stable states, depicted in the 
snapshots of the upper row, which were obtained for $z_B=1.3$ and $L=12$. For 
clarity, only the $A$ particles and obstacles are shown. The first state is a 
homogeneous gas phase, characterized by a low density of $A$ particles (left 
frame). The second is an {\it inhomogeneous} state, with a high (low) density of 
$A$ particles in the lower (upper) half of the box (center frame). The third is 
a homogeneous liquid phase, characterized by a high density of $A$ particles 
(right frame). The graph shows the corresponding OPD at three values of $z_A$. 
For $z_A=z_B$, a single peak is observed at intermediate density (solid curve); 
for $z_A<z_B$, a bimodal distribution is obtained with the second peak appearing 
at low density (dashed curve); for $z_A>z_B$, a bimodal distribution is also 
obtained but with the second peak appearing at high density (dotted curve).}

\end{center}
\end{figure}

We now consider an athermal binary mixture in the presence of quenched disorder. 
We thus have mobile particles $A$ and $B$ as before, but also quenched obstacles 
which are distributed at random locations at the start of each simulation after 
which they remain frozen. The $A$ and $B$ particles then diffuse through the 
quenched environment. From the set of all possible configurations of quenched 
disorder, we consider a finite number of $N$ configurations, and for each of 
them the OPD is measured. Instead of a single OPD as in \eq{eq:opdab}, we thus 
have a set of distributions
\beq
 P_{L,i}(\rho_A|z_B,z_A), 
\eeq
where $i=1,\ldots,N$ labels the different configurations. Even though we are 
interested in the OPDs for different fugacities, it is sufficient to simulate 
only a few combinations: having simulated a system at some $(z_A, z_B)$, the 
OPDs for nearby fugacities can be obtained via histogram reweighting 
\cite{ferrenberg.swendsen:1988}. To facilitate an efficient storage of the data, 
the approach of \cite{citeulike:7690917, citeulike:3523819} is used. Meaningful 
results around the critical point require $N \sim 10^{3-4}$, so the 
computational effort is much larger compared to the pure systems.

\subsection{extracting the free energy barrier}

To obtain the quenched-averaged free energy barrier, we extract the free energy 
barrier $\Delta F_{L,i}$ for each configuration separately, as in \fig{fig1}(a), 
and then average these values. To be precise: for a given $z_B$, each OPD is 
tuned to coexistence via \eq{eq:cx}, which implies that $\zacx$ varies between 
disorder configurations \cite{citeulike:3523819, citeulike:7690917}. By tuning 
the distributions separately, we ensure that the majority of them become 
bimodal, such that a barrier can indeed be \ahum{read-off}. From the 
individual barriers, the quenched-averaged barrier is obtained as a simple 
average:
\beq\label{eq:dF1}
 \Delta F_L = (1/N) \sum_{i=1}^N \Delta F_{L,i}.
\eeq
The finite number of disorder configurations used in calculating the barrier 
gives rise to a statistical deviation
\beq \label{eq:uF1}
 \uF^2 \defas \frac {1}{\sqrt{N(N-1)}} 
 \sum_{i=1}^N ( \Delta F_{L,i} - \Delta F_L )^2
\eeq
which represents the expected deviation of our result from the result we were
to obtain for $N \to \infty$.

In practice, obtaining the barrier $\Delta F_{L,i}$ can be problematic. For some 
disorder configurations, the OPD does not feature a clear liquid and gas peak 
under the criterion of \eq{eq:cx}. In these cases, an intermediate peak is seen, 
at a density roughly between that of the liquid and gas. This extra peak can 
occur as a replacement for either the liquid or gas peak, or as an additional 
third peak (\fig{fig1}(b)). The existence of these \ahum{problematic} 
distributions is due to cavities formed by the obstacles. In some disorder 
configurations, the obstacles create large regions with a local preference for 
either $A$ or $B$ particles. One can easily show that such regions induce an 
intermediate peak (\fig{fig:snapshots}). We emphasize that around the critical 
point such problematic distributions are scarce in the full ensemble of disorder 
configurations (although modified hyperscaling does allow a finite fraction of 
them \cite{citeulike:7690917}).

\subsection{analysis of finite-size effects: scaling plots}
\label{sec:sp}

Near the critical point, we expect $\Delta F_L$ to scale according to 
\eq{eq:df}. Following standard FSS practice \cite{binder.heermann:2002, 
newman.barkema:1999}, we test \eq{eq:df} using scaling plots. That is, we plot 
$y_L \defas L^{-\theta} \Delta F_L$ versus $\tau \defas t L^{1/\nu}$, $t$ given 
by \eq{eq:t}. Then, if the correct values for $\theta$, $\nu$, and $\tc$ are 
used, the curves $y_L(\tau)$ for different system sizes $L$ should collapse on 
top of each other. The practical difficulty with this method is assessing the 
quality of the collapse: seemingly good data collapses are obtained over a range 
of parameter values, and simple \ahum{eye gauging} becomes unreliable. To obtain 
the parameters of best collapse, as well as a measure of the reliability of the 
results, a numerical procedure was therefore used. 

Given a candidate tuple $(\theta, \nu, \tc)$, \eq{eq:uF1} directly yields the 
statistical uncertainty in $y_L(\tau)$, namely $\delta y_L (\tau) = L^{-\theta} 
\uF$. We average $\delta y_L(\tau)$ over the different system sizes to obtain 
the typical statistical uncertainty $u(\tau)$ of the curves at value~$\tau$. For 
a given $\tau$, the curves $y_L(\tau)$ for the different system sizes spread 
around some average value with a root-mean-square width $\sigma(\tau)$. We then 
compare $\sigma(\tau)$ to the typical statistical uncertainty $u(\tau)$ via the 
ratio $\sigma(\tau)/ u(\tau)$. This ratio measures how strong the curves differ 
compared to the random scatter induced by the finite number~$N$ of obstacle 
configurations. To assess the quality of the collapse, we then compute the 
average of the ratio over some region of $\tau$ around the critical point
\beq \label{def:R} 
 R(\theta, \nu, T_c) \defas 
 \frac{1}{\tau_1- \tau_0} \int_{\tau_0}^{\tau_1} 
 \frac{\sigma(\tau)}{u(\tau)} \, d\tau.
\eeq
Values $R<1$ are interpreted as fully consistent with a perfect collapse of the 
curves, while $R>1$ is considered less-consistent. The integration range is 
chosen as large as possible, but small enough to stay within the critical 
region. We used $\tau_0 = -0.2 \cdot 10^{-1/\nu}$ and $\tau_1 = 0.2 \cdot 
10^{-1/\nu}$, which assumes a typical system size $L \sim 10$, and restricts the 
temperature range to $|t|<0.2$.

\section{Results}

\subsection{RFIM}

\begin{figure}
\begin{center}
\includegraphics[width=\figwidth]{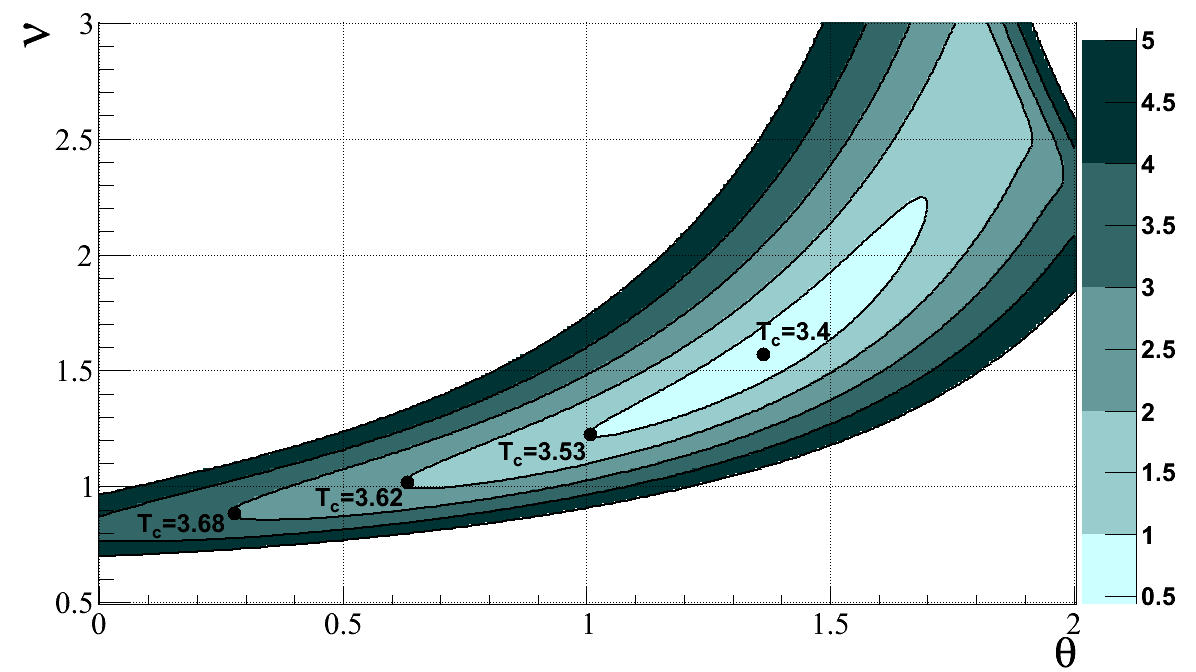}
\caption{\label{fig:RFIMplot} Scaling analysis for the RFIM. Shown are contours 
of $\rmin$ in the plane of critical exponents $(\theta,\nu)$; dots mark the 
critical temperature $\tc$ at a few selected points, the best estimate being 
$\tc \sim 3.4$. The plot shows a clear preference for the expected RFIM value 
$\theta \sim 1.5$, while $\theta=0$ of the pure Ising model, as well as 
$\theta=2$ of a first-order transition, are effectively excluded.}
\end{center}
\end{figure}

To provide a benchmark point for our analysis, we first apply our scaling test 
to data for the RFIM model. These data were taken from our previous work 
\cite{citeulike:7690917}, and obtained for the RFIM in three dimensions using a 
random field drawn from a Gaussian distribution. The total number of disorder 
configurations per system size equals $N \sim 10^4$. For system sizes $L=8$, 
$10$, $14$ and $16$, we examine $R(\theta, \nu, T_c)$ in the plane of critical 
exponents $(\theta, \nu)$. For each point in the plane, the critical temperature 
$\tc$ is tuned such that $R(\theta, \nu, \tc)$ is minimized; the value at the 
minimum is denoted $\rmin$. The results are collected in \fig{fig:RFIMplot} as 
contours. The region where $\rmin \leq 1$, i.e.~where the collapse is fully 
consistent, agrees with the expected value $\theta \sim 1.5$. Also of interest 
is that the region of good collapse features a tail extending toward the pure 3D 
Ising values $(\theta=0, \nu \sim 0.63)$~\cite{citeulike:4844804}. The cause of 
this tail is not entirely clear, but we suspect it is due to crossover effects 
\cite{citeulike:4822813}. However, at the pure Ising values, the deviation 
between the curves $y_L$ exceeds the statistical fluctuation due to the finite 
number of disorder configurations by a factor of more than three. We conclude 
that the pure Ising exponents are effectively excluded by our analysis. Note 
that $\theta=d-1=2$, i.e.~the \ahum{exponent} one would observe below the 
critical temperature (where the transition is first-order) is also excluded.

\subsection{WR model with quenched obstacles}

\begin{figure}
\begin{center}
\includegraphics[width=\figwidth]{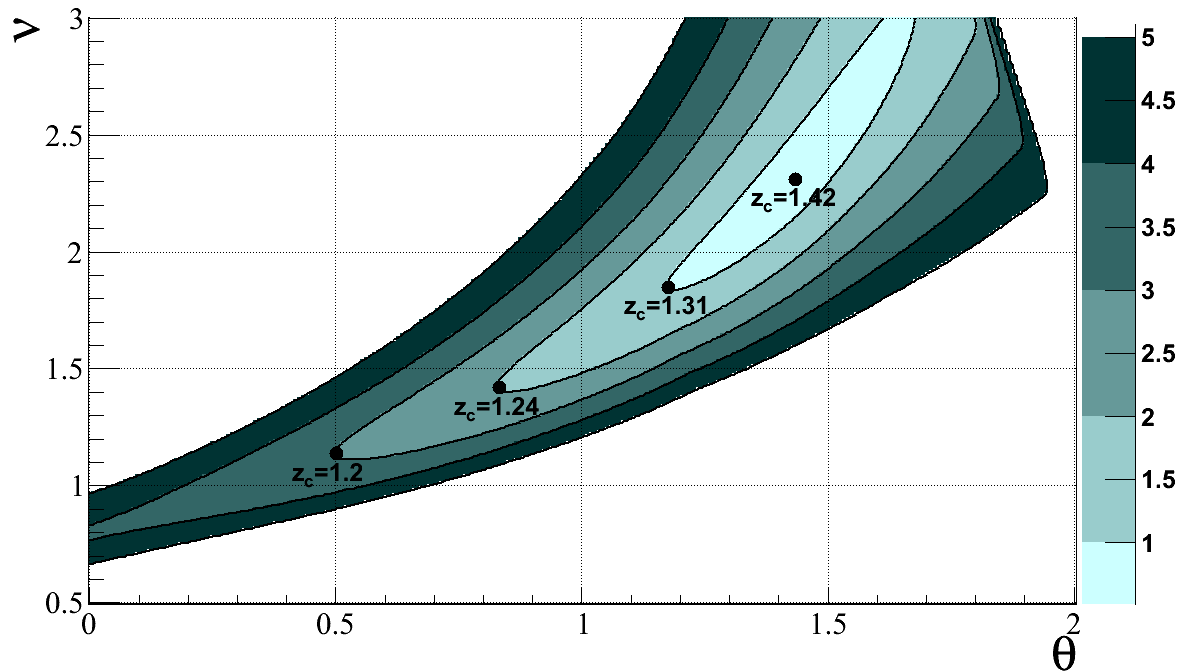} 
\caption{\label{fig:WR1} The analogue of \fig{fig:RFIMplot} for the WR model 
with quenched obstacles; dots indicate the critical fugacity $\zc$ at some 
selected points in the $(\theta,\nu)$ plane. The global minimum of $\rmin$ is at 
$\zc \sim 1.42$, which thus serves as our best-estimate for the critical 
fugacity. As for the RFIM model, a clear preference for $\theta \sim 1.5$ is 
found, as well as the exclusion of $\theta=0$ and $\theta=2$.}
\end{center}
\end{figure}

The WR model \cite{widom.rowlinson:1970} consists of unit-diameter spheres, of 
species $A$ and $B$. The only interaction is a hard-core repulsion between 
particles of the opposite species. We consider the WR model in the presence of 
quenched obstacles; the spatial dimension is $d=3$. The disorder configurations 
are generated by placing $0.02 \times L^d$ particles of species $A$, and the 
same number of species $B$, into a periodic cube of edge $L$ (non-integer 
numbers are appropriately rounded to the next integer at random). The obstacles 
are inserted at random positions, irrespective of overlap between them, after 
which they remain quenched. Next, mobile particles are introduced and these 
interact with the quenched particles following the usual WR rule: quenched 
$A$-obstacles ($B$-obstacles) have a hard-core interaction with mobile $B$ ($A$) 
particles. In this work, system sizes $L=7, 8, \dots, 12$ are used. For each 
system size, OPDs for $N \sim 10^4$ disorder configurations were generated at 
fugacities $z_A = z_B = 1.4$, which is the approximate location of the critical 
point as anticipated from our previous simulations \cite{citeulike:7690917}; all 
details regarding the simulation method can also be found in that reference.

To obtain $\Delta F_L$ via \eq{eq:dF1} over a range of $z_B$ requires extracting 
the free energy barrier from several million(!) distributions, which obviously 
cannot be done by hand. To deal with the \ahum{problematic} distributions of 
\fig{fig1}(b) automatically, a numerical filter was therefore applied: if one of 
the two dominating peaks in the OPD is centered in the intermediate density 
range $0.20 z_B < \rho_A < 0.75 z_B$, the distribution is considered to not 
feature a distinct liquid and gas peak. For these distributions we set $\Delta 
F_{L,i}=0$. Since for the values of $z_B$ considered here no more than $\sim 0.1 
\%$ of the distributions failed this criterion, our results should not 
significantly depend on the filter. The results for $\rmin$, as well as some 
estimates for the critical fugacity $\zc$ obtained from minimizing $R$, are 
shown in \fig{fig:WR1}. Note that, for the WR model, the analogue of \eq{eq:t} 
becomes $t \defas \zc / z_B-1$. As in the RFIM, \fig{fig:WR1} features a clear 
preference for $\theta \sim 1.5$, and the competing values $\theta=0$ and 
$\theta=2$ of, respectively, the pure Ising model and a first-order transition, 
are effectively excluded. We interpret this finding as a strong sign that the WR 
model with quenched obstacles indeed belongs to the universality class of the 
RFIM. Note, however, that we cannot provide high-precision estimates of the 
critical exponents, nor of the critical \ahum{inverse temperature} $\zc$. There 
is a substantial range of possible values $(\theta, \nu, \zc)$ in the region 
$\rmin \leq 2$, i.e.~the region where our results for the WR model and RFIM 
become quantitatively compatible.

\subsection{Colloid-polymer mixtures in porous media}

We now consider a colloid-polymer mixture confined to a porous medium. The 
colloids and polymers are unit-diameter spheres, with only a hard-core repulsion 
between colloid-colloid and colloid-polymer pairs (this is just the AO model 
\cite{asakura.oosawa:1954, vrij:1976} with equally sized colloids and polymers). 
As porous medium we use a quenched configuration of polymers, which are 
distributed randomly in a periodic cube of edge $L$ at the start of each 
simulation run (the average packing fraction of the quenched configuration 
$\eta_Q=0.05$). The OPD for the AO model is the probability distribution 
$P_{L,i}(\eta_{\rm col} | z_{\rm col}, \etapr)$ \cite{vink.horbach:2004*1}, 
where $i$ denotes the quenched polymer configuration for which the OPD was 
measured, $\eta_{\rm col}$ the colloid packing fraction, $z_{\rm col}$ the 
colloid fugacity, and $\etapr$ the polymer reservoir packing fraction. Note 
that, for the AO model, $\etapr$ is just the polymer fugacity multiplied by the 
volume of a single polymer, and thus plays the role of inverse temperature. 
Consequently, the analogue of \eq{eq:t} becomes $t \defas \etapr / \etaprcr - 
1$. One of us (RV) recently studied the above model using the same type 
of quenched disorder. We now extend the analysis of that previous work 
\cite{citeulike:3523819} to the scaling of the free energy barrier.

\begin{figure}
\begin{center}
\includegraphics[width=0.75\columnwidth]{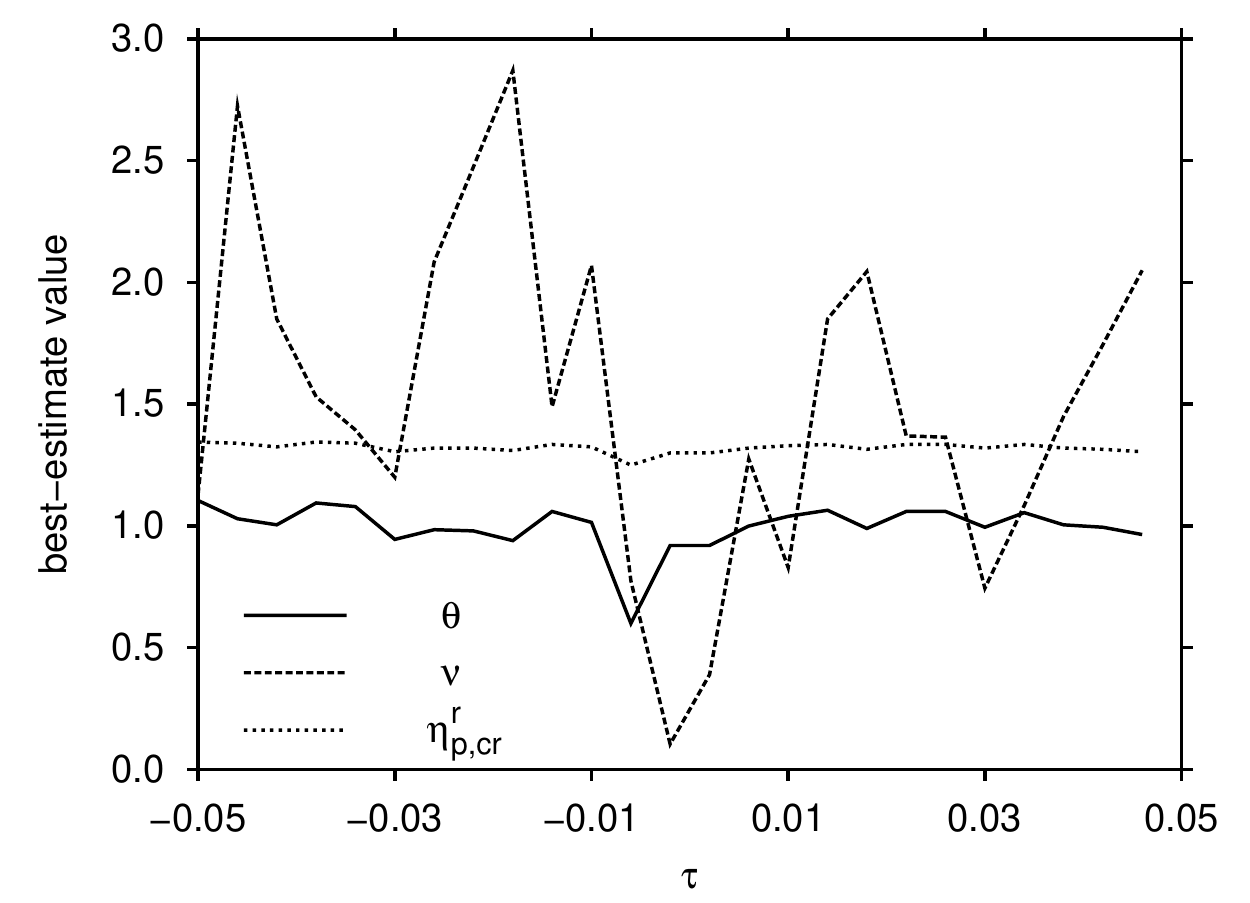}
\caption{\label{fig:ao} Best-estimates of $(\theta, \nu, \etaprcr)$ versus the 
scaling variable $\tau$ for the AO model inside a porous medium; see details in 
text.}
\end{center}
\end{figure}

One practical problem is that the number of disorder configurations in our AO 
data set is \ahum{only} $N \sim 10^3$, i.e.~one order of magnitude below the 
value used in the analysis of the RFIM and WR model. Also the quality of 
individual distributions was worse compared to that of the latter two models. A 
further complication is that the AO model is asymmetric: the coexistence colloid 
fugacity is not trivially related to $\etapr$. In contrast, for the WR model, it 
holds that $z_A=z_B$ at the critical point due to symmetry. Because of all these 
problems, the construction of contour plots, such as \fig{fig:RFIMplot} and 
\fig{fig:WR1}, was not feasible for the AO model, and so a modified analysis is 
presented. To this end, we extract the barrier $\Delta F_L$ from the 
quenched-averaged OPD, following the method outlined in the Appendix of 
Ref.~\cite{citeulike:7690917}. This procedure yields a result similar to 
\eq{eq:dF1}, but it is less susceptible to statistical uncertainties in 
individual distributions. Since the barrier is extracted from the 
quenched-averaged OPD, we loose information about the fluctuations between 
disorder configurations. In particular: \eq{eq:uF1} can no longer be used to 
calculate the typical statistical uncertainty $u(\tau)$ defined in 
\sect{sec:sp}. Hence, in the construction of the scaling plot $y_L(\tau)$, we 
cannot compare the deviation between the curves $\sigma(\tau)$ to the 
statistical uncertainty $u(\tau)$. Only best-estimates can be provided, which are 
obtained by minimizing the relative deviations between the curves.
That is, for a given $\tau$ we minimize the value $\sigma(\tau)/y(\tau)$, where
$y(\tau)$ is the average value of the $y_L(\tau)$.
For $-0.05 < \tau < 0.05$, and system sizes $L=10,11,12$ we computed these 
best-estimates: their variation with $\tau$ is shown in \fig{fig:ao}. If the 
collapse was perfect, the best-estimates should not depend on $\tau$. Regarding 
$\etaprcr$ and $\theta$ this is confirmed, and from the average of the curves we 
conclude that $\etaprcr \sim 1.3$ and $\theta \sim 1.0$ for the AO model. 
However, it is clear from \fig{fig:ao} that no meaningful estimate of $\nu$ can 
be provided. We emphasize that $\etaprcr \sim 1.3$ deviates by about 10\% from 
previous estimates. The reason is that, in Ref.~\cite{citeulike:3523819}, we 
assumed $\nu_{\rm RFIM}=1.1$. However, from the analysis of the RFIM 
(\fig{fig:RFIMplot}) it transpires that $\nu_{\rm RFIM}$ may well be different. 
This is furthermore corroborated by the literature \cite{nattermann:1998}, where 
large variations in $\nu_{\rm RFIM}$ are also reported.

\section{Conclusions}	
\label{con}

The aim of this paper was to test the relation $\Delta F_L \propto L^\theta$ in 
fluids with quenched disorder. This relation applies when the universality class 
is that of the RFIM \cite{citeulike:7690917}. For the WR and the AO model with 
quenched obstacles, our data confirm that $\Delta F_L$ diverges as a power law 
with system size at the critical point. This is an important qualitative 
indication that the universality class is no longer that of the pure Ising 
model, since then $\Delta F_L$ would be constant at criticality. For the WR 
model, $\theta$ is consistent with the RFIM value $\theta_{\rm RFIM} \sim 1.5$. 
For the AO model, $\theta \sim 1.0$ is found, which is somewhat below the RFIM 
value. Possible reasons for this discrepancy are (1) crossover effects 
\cite{citeulike:4822813} (in this case from pure Ising to RFIM universality), 
(2) corrections to scaling due to asymmetry, and (3) insufficient disorder 
averaging. We believe that the crossover scenario is the most likely 
explanation, since $\theta \sim 1.0$ is \ahum{in-between} the pure Ising value 
$\theta=0$ and the RFIM value. Of course, all these problems could be solved by 
simulating larger systems and more disorder configurations, but the cost in CPU 
time is still prohibitively large. Our data also make clear that high-precision 
estimates of critical exponents and temperatures in random-field systems remain 
difficult to obtain in simulations. The data for the RFIM and the WR model 
reveal large variations, see \fig{fig:RFIMplot} and \fig{fig:WR1}, in particular 
for the correlation length critical exponent $\nu$.

\ack
This work was supported by the {\it Deutsche Forschungsgemeinschaft} under the 
Emmy Noether program (VI 483/1-1).

\section*{References}
\bibliography{refs}

\end{document}